\documentclass[runningheads,a4paper]{llncs}

\usepackage{amssymb}
\usepackage{graphicx}

\usepackage[noend]{algpseudocode}
\usepackage{algorithm}

\usepackage{url}
\urldef{\mailsa}\path|{marek.kokot, sebastian.deorowicz,agnieszka.grabysz}@polsl.pl |
\urldef{\mailsb}\path|agnieszka.grabysz, bozena.wieczorek}@polsl.pl|

\newcommand{\keywords}[1]{\par\addvspace\baselineskip
\noindent\keywordname\enspace\ignorespaces#1}

\begin{document}

\mainmatter  

\title{Sorting Data on Ultra-Large Scale with RADULS\\New Incarnation of Radix Sort}

\titlerunning{New Incarnation of Radix Sort}

%

%
%
\author{Marek Kokot\and Sebastian Deorowicz\and Agnieszka Debudaj-Grabysz
}
\authorrunning{M.\ Kokot, S.\ Deorowicz, A.\ Debudaj-Grabysz, B.\ Wieczorek}

\institute{Institute of Informatics, Silesian University of Technology, Akademicka 16,\\ 44-100 Gliwice, Poland\\
\mailsa
}

%
%

\maketitle

\begin{abstract}
The paper introduces RADULS, a new parallel sorter based on radix sort algorithm, intended to organize ultra-large data sets efficiently.
For example 4\,G 16-byte records can be sorted with 16 threads in less than 15 seconds on Intel Xeon-based workstation. 
The implementation of RADULS is not only highly optimized to gain such an excellent performance, but also parallelized in a cache friendly manner to make the most of modern multicore architectures.
Besides, our parallel scheduler launches a few different procedures at runtime, according to the current parameters of the execution, for proper workload management.
All experiments show RADULS to be superior to competing algorithms. 
\end{abstract}

\section{Introduction}
Although the area of sorting algorithms has been investigated 
from time immemorial, there is still the need for developing faster implementations as well as the place for improvement. 
The demand for even faster sorters is the result of accumulation of increasing amounts of data in all areas of life. 
Starting from user applications, through industry, to strictly scientific applications, organizing the data is required everywhere.
It seems that the possibility of constructing 
{\it new} sorting algorithms in the strict sense has been exhausted, that is why faster sorters are constructed by using the following techniques:\
\begin{itemize}
\item widening the range of applications of internal sorting,
\item parallelization,
\item hardware friendliness, 
\item hybrids of the above mentioned.
\end{itemize}

The range of applications of internal sorting algorithms become wider naturally, due to the availability of increasingly larger memory sizes.
At the same time, the prevalence of multicore architectures gives the possibility of using thread-level parallelism to increase computing power.
The programming environments that supports developing of multithreaded programs become more and more popular.
Unfortunately, the effort in the parallelization of a code could be of no effect if the algorithm were not architecture friendly, especially cache friendly.
The subject is crucial for single-threaded algorithms, but becomes even more critical for parallel ones.

Concerning modern architectures it is known, that the performance bottleneck is accessing the main memory. 
The problem is being solved by equipping CPUs with cache systems to reduce the time to access data.
Most often cache memories have a hierarchical structure. 
When accessing data a processor tries to find it in the first-level (L1) cache,   then in case of a fail---at higher levels (L2, L3).
If the data was not stored in the cache, it is necessary to tap into the main memory with much longer latency.
This is called {\it a cache miss}.

For multicore architectures individual cores have their own, private first-level caches, while last-level cache is typically shared.
Memory access policy have to coordinate several problems, e.g., keeping coherent data of private caches, controlling shared data accesses, especially in case of cache misses.
As caches are organized in lines, there is a conflict when separate threads residuing on separate cores request access to separate data for modification, but the data falls in the same cache line.
This is called {\it false sharing}~\cite{ref:Int2011}. 
In such a situation a synchronization protocol forces unnecessary memory update to keep coherency.

The sorting algorithm which is very well suited for parallelization is radix sort.
It represents non-comparison based sorts~\cite{ref:Sed1998} with the low computational complexity of $O(N)$, where $N$ is the number of elements to sort.
Keys to be sorted are viewed as sequences of fixed-sized pieces, e.g., decimal numbers are treated as sequences of digits while binary numbers are treated as sequences of bits.
Generally these pieces correspond to {\it digits} that creates numbers represented in a base-$R$ number system, where $R$ is its {\it radix}.

Still, there are two basic approaches to radix sort. 
According to the first variant, digits of the keys are examined from the most to the least significant ones (MSD).
In the second variant, digits are processed in the opposite direction, i.e. from the least to the most significant ones (LSD).
The general idea of the algorithm is that at each radix pass the array of keys is sorted according to every consecutive digit.
The number of passes corresponds to the length of a key.  
The most often a counting sort is selected as the inner sorter.
In case of LSD sorting, it is not intuitive, that the final distribution is properly ordered. 
In fact, the usage of a stable method like the counting sort (preserving the relative order of items with duplicated keys in the array), guaranties accuracy of results.
Two traversals through the array are required at each pass.
During the first one, a histogram of the number of occurrences of each possible digit value is obtained.
Next, for each key, the number of keys with smaller (or the same) values of digits on investigated position is calculated. 
Finally the keys are distributed to their appropriate positions on the basis of the histogram, during the second traversal through the array.
The idea of MSD method is the same, although it is worth to mention, that after the first pass of sorting, the array of keys is partitioned into a maximum of $R$ different {\it bins}. 
Every single bin contains keys for one possible value of the most significant byte.
During every consecutive pass the keys are sorted within bins from the previous pass and they are not distributed among separate bins.


In this paper we propose RADULS, which is a parallel radix sorter capable to sort 4\,G 16-byte records in less than 15 seconds using 16-cores.
A few parallelization strategies are cooperating to assure load balancing. 
Additionally, RADULS is cache friendly to reduce long latencies of accesses to the main memory.

The paper is organized as follows.
In Section~\ref{sec:related} a brief description of the state-of-the-art radix sorters is given.
Section~\ref{sec:algorithm} describes our algorithm, which is experimentally evaluated against the top parallel sorters (radix-based and comparison-based) in Section~\ref{sec:experiments}.
In Section~\ref{sec:applications} we discuss some of the applications of radix sort.
The last section concludes the paper.


\section{Related Works}
\label{sec:related}
The area of sorters using thread-level parallelism was broadly investigated in recent decades.
Our inspiration was the paper by Satish \emph{et al.}~\cite{ref:Sat2010}, where, among others, an architecture-friendly LSD radix sorter for CPU was proposed. 
Firstly, the authors identified bottlenecks in common, parallel implementations, such as irregular memory accesses or conflicts in cache or shared memory accesses.
Secondly, they proposed the ways to avoid them.
The idea is to maintain buffers in local storages for collecting elements, that belong to the same radix. 
The buffers for $B \times R$ bytes in local memory for each thread is reserved, where $R$ is the radix, and $B$ bytes are buffered for each radix.
The $B$ value was selected as a multiple of 64 bytes (cache line size).
Buffers' contents are written to global memory when enough elements were accumulated and then is reused for other entries.
Such an approach avoids cache misses, as buffers occupy contiguous space and fit the cache memory.

In~\cite{ref:Cho2015} an in-place parallel MSD radix sorter was proposed. 
There is no auxiliary array available in the distribution phase, that is why swap operations are performed in order to place keys to their proper bins. 
This phase is especially challenging when parallelizing, because of existing dependencies between reading and writing within swaps, while the array of keys is  partitioned among threads.
Hence, the authors solve the problem in two stages called {\it speculative permutation} and {\it repair}.
The stages are iterated until all the keys are rearranged and placed in their target bins.
Next, the bins can be sorted independently. 
However, these sub-tasks can be heavily imbalanced, because of differences in bin sizes. 
That is why the authors use adaptive thread reallocation scheme to gain proper load balancing.

\section{Our Algorithm}
\label{sec:algorithm}

\subsection{General idea}
The algorithm follows the MSD approach with radix $R=256$.
Our parallel scheduler launches four different procedures for single radix pass depending on the current digit position and the size of the bin.
It also uses some special treatment of large bins for better balancing of work among threads.
Finally, sufficiently small bins are handled by comparison-based sorting routines.
In the following subsections we describe each of the procedures in details.

%
%

\subsection{First-digit pass} \label{sec:first_pass}
Figure~\ref{alg:first_pass} presents a general scheme of sorting according to the first digit.
At the beginning the keys are distributed into bins by $T$ threads using buffered radix split algorithm inspired by a single pass of Satish~\emph{et al.}~\cite{ref:Sat2010} LSD radix sort.
For better load balancing we initially divide the input array into $8T$ chunks of sizes linearly growing from $N / 64T$, where $N$ is the array size (constants chosen experimentally).

 
\begin{figure}[h]
	\begin{algorithmic}[1]
	\Function{FirstPass}{$\mathit{data\_start}, \mathit{data\_end}, \mathit{current\_byte}, T$}
			\State $bins \leftarrow \Call{BufferedRadixSplit}{\mathit{data\_start}, \mathit{data\_end}, \mathit{current\_byte}}$
		\If{$\mathit{current\_byte} > 0$}
		\State $[\mathit{small\_bins}, \mathit{big\_bins}] \leftarrow \Call{SplitBins}{\mathit{bins}}$
		\ForAll{$\mathit{bin}$ in $\mathit{small\_bins}$}
		\State $\Call{task\_queue.put}{\mathit{bin}, \mathit{current\_byte} - 1}$
		\EndFor		
		\State $T_\mathrm{big} \leftarrow \min(T, 1.25 \times N_\mathrm{big} / N)$ \Comment{$N_\mathrm{big}$---no.\ of rec.\ in all big bins}
		\State $T_\mathrm{small} \leftarrow T - T_\mathrm{b}$
		\For{$i=1$ to $T_\mathrm{small}$} 
		\State Run \Call{ProcessBins}{} in a new thread 
		\EndFor
		\ForAll{$\mathit{bin}$ in $\mathit{big\_bins}$}
		\State $\Call{BigPartitonPass}{\mathit{bin.start}, \mathit{bin.end}, \mathit{current\_byte} - 1, T_\mathrm{big}}$
		\EndFor
		\For{$i=1$ to $T_\mathrm{big}$} 
		\State Run \Call{ProcessBins()}{} in a new thread 
		\EndFor
		\EndIf
	\EndFunction
	\end{algorithmic}
	\caption{Pseudocode of the first pass of RADIUS algorithm}
\label{alg:first_pass}
\end{figure}

When the distribution is over, the bins are marked as small or big.
The threshold is set to $2N / 3T$. 
The idea behind this is to assign sufficiently large number of threads for handling big bins, i.e. to avoid the situation in which threads assigned to processing of small bins completed their work, when the big bin are still processed.
Small bins are intended for processing by a different procedure than big ones.
That is why the priority queue is created for keeping tasks describing small bins.
Tasks are ordered from the largest size of a small bin to the lowest one.
$T_\mathrm{small}$ of newly created threads are to handle the mentioned queue.
Furthermore, when big-bin processing is over, $T_\mathrm{big}$ threads for small-bin processing are created (see Figure~\ref{alg:first_pass}) to replace the released ones.

\subsection{Next passes for big bins} \label{big_bins}
The pass processing big bins (Figure~\ref{Alg:big_bins}) is quite similar to the first-digit pass.
The only difference is that after marking bins as big and small the first of them are processed immediately in a recursive manner, while the later are added to the priority queue. 
The queue of small bins from discussed passes is the separate one.
This implies it is handled only with the threads previously assigned to big bins (after the first pass).
%


\begin{figure}[t]
	\begin{algorithmic}[1]
		\Function{BigPartitonPass}{$\mathit{data\_start}, \mathit{data\_end}, \mathit{current\_byte}, T$}
		\State $\mathit{bins} \leftarrow \Call{BufferedRadixSplit}{\mathit{data\_start}, \mathit{data\_end}, \mathit{current\_byte}}$
		\If{$\mathit{current\_byte} > 0$}
		\ForAll{$\mathit{bin}$ in $\mathit{bins}$}
		\If{$\Call{IsBig}{\mathit{bin}}$}
		\State $\Call{BigPartitonPass}{\mathit{bin.start}, \mathit{bin.end}, \mathit{current\_byte} - 1, T}$
		\Else
		\State $\Call{task\_queue.put}{\mathit{bin}, \mathit{current\_byte} - 1}$
		\EndIf
		\EndFor
		\For{$i=1$ to $T$} 
		\State Run \Call{ProcessBins}{} in a new thread 
	\EndFor
		\EndIf
	\EndFunction
	\end{algorithmic}
	\caption{Pseudocode of the algorithm handling big bins produced in the first stage}
	\label{Alg:big_bins}
\end{figure}

\subsection{Next phases for small bins} \label{normal_bins}
Descriptions of small bins generated in previous passes are kept in the form of tasks in a priority queue.
Each available thread processes these tasks one by one. 
Figure~\ref{Alg:normal_bins} shows the pseudocode of a single-threaded algorithm handling the tasks from the queue.

To sort a bin according to some digit one of following methods is chosen relating on the size of the bin.
Simple counting sort without buffering the keys is used when the bin fits a half of L2 cache.
In the opposite case, the buffering algorithm inspired by single pass of Satish~\emph{et al.} algorithm  is used. 
The case of tiny bins is discussed in the following subsection.

\begin{figure}[h]
	\begin{algorithmic}[1]
		\Function{MSDRadixBins}{$\mathit{data\_start}, \mathit{data\_end}, \mathit{current\_byte}$}
		\If{data fit in cache}
			\State $\mathit{bins} \leftarrow \Call{RadixSplit}{\mathit{data\_start}, \mathit{data\_end}, \mathit{current\_byte}}$
		\Else
			\State $\mathit{bins} \leftarrow \Call{BufferedRadixSplit}{\mathit{data\_start}, \mathit{data\_end}, \mathit{current\_byte}}$
		\EndIf
		\If{$\mathit{current\_byte} > 0$}
		\ForAll{$\mathit{bin}$ in $\mathit{bins}$}
		\If{$\Call{IsTinyBin}{\mathit{bin}}$}
		\State $\Call{ComparisonSort}{\mathit{bin}}$
		\Else
		\If{$\Call{tooSmallForQueue}{\mathit{bin}}$}
			\State $\Call{MSDRadixBin}{\mathit{bin.start}, \mathit{bin.end}, \mathit{current\_byte} - 1}$
		\Else
			\State $\Call{task\_queue.put}{\mathit{bin}, \mathit{current\_byte} - 1}$
		\EndIf
		\EndIf
		\EndFor
		\EndIf
	\EndFunction
	\Function{ProcessBins}{}
	\While{$[\mathit{bin}, \mathit{byte}] \leftarrow \Call{tasks\_queue.pop}{ }$}
	\State $\Call{MSDRadixBins}{\mathit{bin.start}, \mathit{bin.end}, \mathit{byte}}$
	\EndWhile
	\EndFunction
	\end{algorithmic}
	\caption{Pseudocode of the algorithm handling small bins produced in previous stages}
	\label{Alg:normal_bins}
\end{figure}

The new bins obtained in this place can be handled in three ways.
If the bin size is smaller than $N/4096$, it is processed recursively to avoid too many (potentially costly) operations on the queue (which is shared by many threads).
Larger bins are inserted into the priority queue.


\subsection{Handling tiny bins}
The bins containing smaller than 384 keys (value chosen experimentally) are processed by comparison sorters to avoid relatively costly passes of radix sort.
We experimented with several comparison algorithms, but finally picked three of them: introspective~\cite{ref:Mus1997} (implemented as part of the standard C++ library), Shell sort~\cite{ref:She1959} (with sequence of increments reduced only to $1, 8$), and insertion sort~\cite{ref:Knu1968}.
For the smallest arrays ($N\leq 32$) we use insertion sort.
The threshold between introspective sort (a hybrid of quick sort~\cite{ref:Hoa1962} and heap sort~\cite{ref:Wil1964}) and Shell sort depends on the key size (expressed in bytes), but usually is in the range 100--180.

While deciding whether the current bin is tiny or not we also monitor the ``narrowing factor'' defined as the number of keys in the ``parent'' bin divided by the number of keys in the current bin.
If this factor is larger than $\sqrt{R} = 16$ we speculate that the next pass of the radix sort should be more profitable than using introspective or Shell sort.
Thus in such a situation the tiny bin threshold is set to 32.


\section{Experimental results}
\label{sec:experiments}
RADULS was implemented in the C++14 programming language and uses native C++ threads.
A few SSE2 instructions were used for fast transfers of buffered memory to the main memory without cache pollution.
For compilation we used GCC 6.2.0.
All experiments were performed at workstation equipped with two Intel Xeon E5-2670\,v3 CPUs (12 cores each, 2.3\,GHz) and 128\,GB RAM.

We compared RADULUS with the following parallel sorting algorithms:
\begin{itemize}
\item TBB---the parallel comparison sort of $O(N\log N)$ average time complexity implemented in the Intel Threading Building Blocks~\cite{ref:TBB} (2017 Update 3 release),
\item MCSTL---the parallel hybrid sort~\cite{ref:Sin2007,ref:mcstl}, now included in GNU's libstdc++ library,
\item Satish-1---our implementation of the buffered LSD radix sort introduced by Satish \emph{et al.}~\cite{ref:Sat2010} with the buffer size for a specific digit equal to the cache line size ($B=64$),
\item Satish-4---the same as Saitsh-1, but with $B=256$,
\item PARADIS---the state-of-the-art in-place radix sort algorithm by Cho \emph{et al.}~\cite{ref:Cho2015}. 
\end{itemize}
Unfortunately, we were not able to obtain either the PARADIS source codes or library delivering it.
Since the algorithm is far from being trivial to implement we decided to include in our comparison the running times just from the PARADIS paper without any time scaling (although PARADIS was evaluated at Intel Xeon E7-8837 CPU clocked at higher rate, i.e. 2.67\,GHz).

In the first set of experiments we compared the running times of sorting algorithms for array sizes in the range from 62.5\,M to 4\,G records.
The records of length 16 bytes consisted of 8-bytes-key and 8-bytes-data fields (to allow indirect comparison with PARADIS).
The keys were produced randomly with: uniform distribution and Zipf distribution~\cite{ref:Gra1994} with $\theta=0.75$ (once again to allow comparison with PARADIS results).

\begin{figure}[t]
\begin{center}
\begin{tabular}{ccc}
\includegraphics[width=0.48\textwidth]{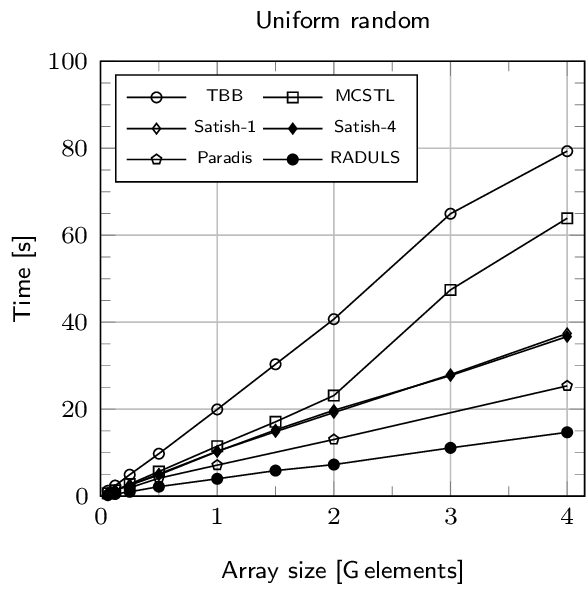} &&
\includegraphics[width=0.48\textwidth]{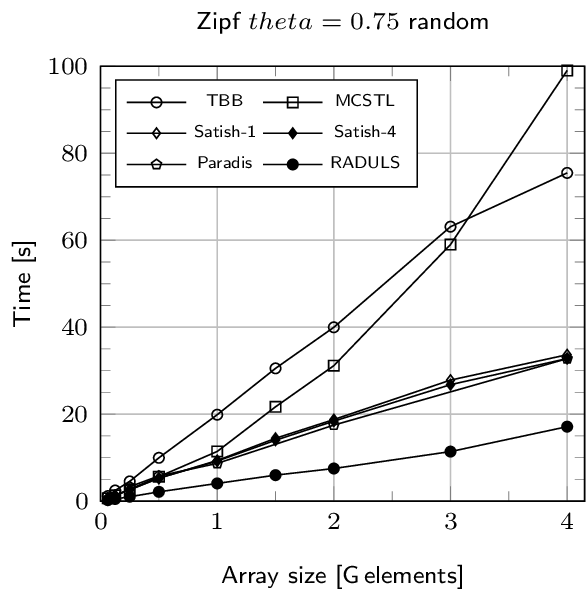}
\end{tabular}
\end{center}
\caption{Experimental results for random input: uniform distribution (left) and Zipf $\theta=0.75$ distribution (right).
The sorted records are of size 16 bytes (8 bytes for key and 8 bytes for data).
The number of threads was set to 16.}
\label{fig:sizes}
\end{figure}

Figure~\ref{fig:sizes} shows that RADULS clearly outperforms the competitive algorithms when run for 16 threads.
PARADIS was the second best for uniform distribution.
The second place for Zipf data are, however, shared by PARADIS and Satish algorithms.
The difference between Satish-1 and Satish-4 is marginal in both cases.

\begin{figure}[th]
\begin{center}
\begin{tabular}{ccc}
\includegraphics[width=0.48\textwidth]{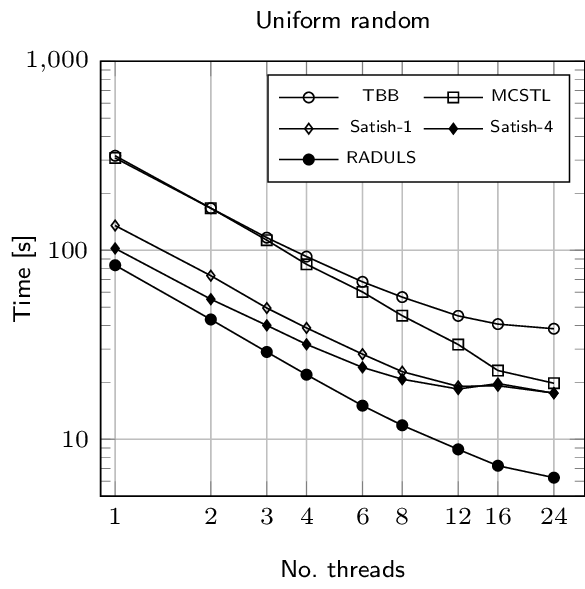} &&
\includegraphics[width=0.48\textwidth]{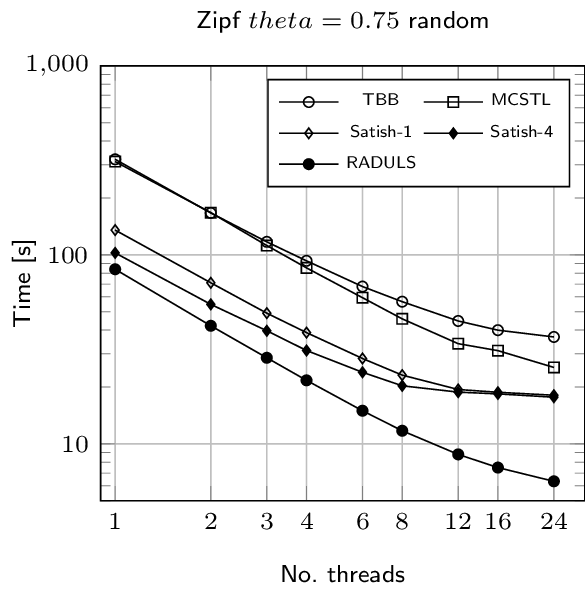}	\\
\includegraphics[width=0.48\textwidth]{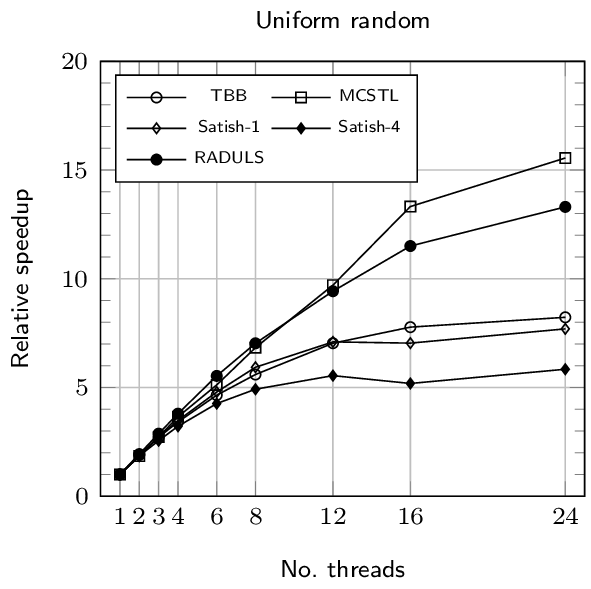} &&
\includegraphics[width=0.48\textwidth]{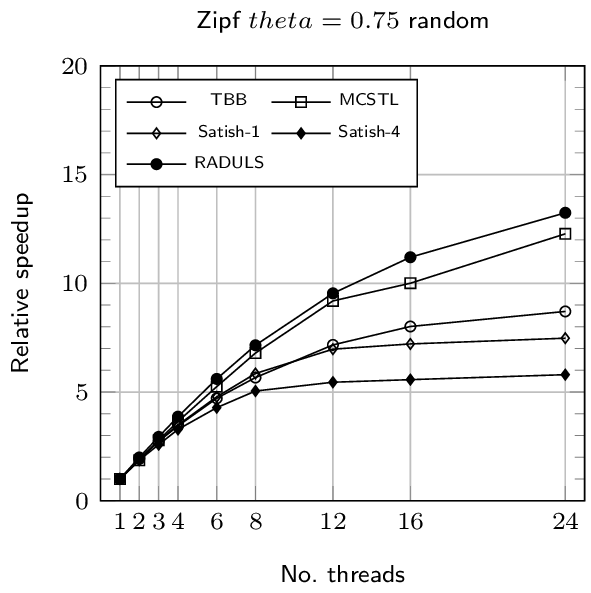}
\end{tabular}
\end{center}
\caption{Experimental results for random input: uniform distribution (left) and Zipf $\theta=0.75$ distribution (right).
The array contained 2\,G elements of size 16 bytes (8~bytes for key and 8 bytes for data).
The number of threads was: 1, 2, 3, 4, 6, 8, 12, 24.}
\label{fig:threads}
\end{figure}

In the second experiment, we evaluated the influence of number of running threads.
Figure~\ref{fig:threads} shows both the absolute running times and the relative speedup of the algorithms.
As it can be observed, the relative speedups of Satish-1 is better than of Satish-4, but the later is faster for smaller number of threads, especially for a single thread.
Both Satish algorithms and TBB scales well only for less than 8 threads. 
Then their speedups saturates below 9.
RADULS scales better and for 16 threads the relative speedup is about 11.5.
MCSTL performs even better (in term of scalability) for uniformly distributed keys.
Nevertheless, its absolute running times are much longer than RADULS times.
The inspection of PARADIS paper (Figure 7c) shows that for 16 threads its speedups is almost~10.

\begin{figure}[t]
\begin{center}
\includegraphics[width=\textwidth]{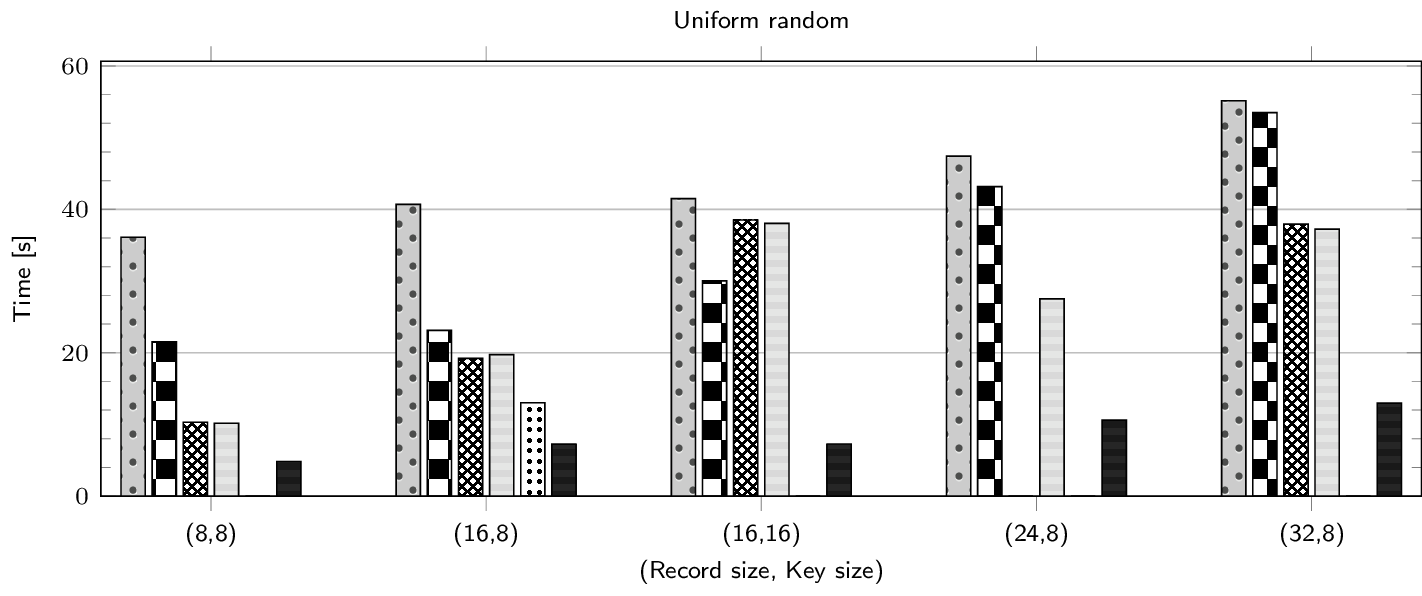}\\
\includegraphics[width=\textwidth]{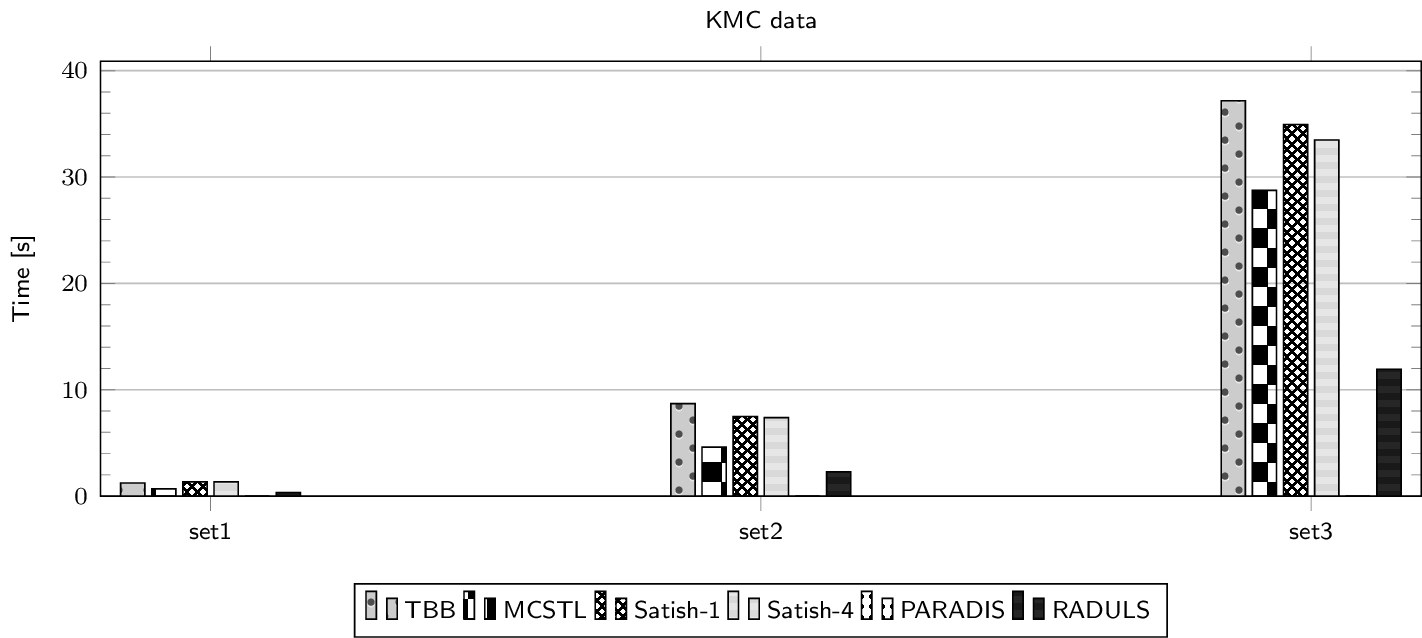}
\end{center}
\caption{Experimental results for records of various size (up) and type (down).
The sets of KMC data contained 16-byte records with 16-byte keys of sizes
63\,M, 413\,M, 1887\,M.
}
\label{fig:records_KMC}
\end{figure}

Finally, we experimented with various record sizes and types of data.
The upper chart in Figure~\ref{fig:records_KMC} shows the running times for records from 8 to 32 bytes with 8-byte (or 16-byte in one case) keys.
It can be noticed that RADULS is always the fastest.
The running times grows from 4.82\,s to 12.97\,s when key size is 8-bytes long and the record size grows from 8 to 32 bytes.
The sublinear time increase was possible due to use of comparison sorting routines for tiny bins.
The lower chart in Figure~\ref{fig:records_KMC} presents the results for three sets of $k$-mers, being the data from large sequencing project (see Section~\ref{sec:applications} for details). 
Once again RADULS appeared to be the winner.

\section{Possible Applications}
\label{sec:applications}
RADULS is a general purpose sorter, but the two of its possible applications are especially worth mentioning. 
We have already tested the first one, as it consisted in joining our sorter with existing $k$-mer counter software, i.e.  KMC~\cite{ref:Deorowicz2013,ref:Deorowicz2015}.
By $k$-mers we mean unique substrings of length $k$ in a set of reads from sequencing projects.
The procedure of determining $k$-mers is often used in initial stages of sequencing data processing.
The input data can be larger than 1\,TB.
Therefore, modern $k$-mer counters usually process in two stages.
In the first stage the extracted $k$-mers are distributed into several hundred disk files, which are then processed separately (the second stage).
One of the possibilities of handling a single file is to sort the strings and then remove duplicates.
In fact, this solution was used in KMC.

Another application of the proposed algorithm could be to sort short strings.
Let's consider searching multiple patterns in a text with the aid of a suffix array, a classical full-text index applying a binary search of a pattern against a collection of sorted suffixes of the text.
It could be performed in two ways.
In a naive solution each pattern is sought separately.
However, a better way could be to sort the patterns (or their prefixes only) first, and then to search them in the suffix array in an incremental manner, i.e. with a reduced range for the binary search, which may be faster overall.
It is vital to use an efficient sorter in the preliminary phase of this procedure.

\section{Conclusions}
Although the art of sorting algorithms seems to be thoroughly understood, technological progress and new development tools allow the creation of more and more efficient sorters.
In out paper we propose radix-based sorter, which owes its outstanding performance to an innovative combination of several techniques.
It is parallelized in a cache friendly manner---thus adapted to modern multicore architectures.
Due to maintaining software-managed buffers for collecting data, which are to be flushed to the main memory, long latencies are reduced.
The parallel scheduler---being a part of out software---allows sub-task-size-driven execution to avoid workload imbalance.
On the basis of monitoring current parameters an appropriate number of threads per sub-task and a proper sorting method can be selected.
Beside of radix algorithm, introspective, Shell and insertion sort algorithms are incorporated.
Finally, RADULS is highly optimized using the latest advances in software compilers.
%
%
Experiments show that RADULS outperforms its competitors for both uniformly distributed data as well as for skewed one.
That implies it may become an irreplaceable sorter for a wide range of applications.

\subsubsection*{Acknowledgments.}
The work was supported by the Polish
National Science Centre under the project DEC-2013/09/B/ST6/03117.


\bibliographystyle{splncs03}
\bibliography{ref_1284}

%
%
%
%
%
\end{document}